%% file: paper.tex
\documentclass{llncs}
\usepackage{graphicx}
\usepackage{xspace}
\usepackage{url}
\makeatletter
\def\@begintheorem#1#2{\sl \trivlist \item[\hskip \labelsep{\bf #1\ #2:}]}
\def\@opargbegintheorem#1#2#3{\sl \trivlist
      \item[\hskip \labelsep{\bf #1\ #2\ #3:}]}
      \makeatother

\newcommand{\R}{{\mbox{\textbf{R}}}}

\begin{document}
\title{\LARGE Choosing Colors for Geometric Graphs
              via Color Space Embeddings}
\date{}
\author{
{Michael B. Dillencourt}
\and
{David Eppstein}
\and
{Michael T. Goodrich}
}
\institute{Dept.~of Computer Science, 
    Univ.~of California, Irvine, CA 92697-3425 USA.
    \email{\{dillenco,eppstein,goodrich\}(at)ics.uci.edu.}}
\maketitle

\begin{abstract}
Graph drawing research traditionally focuses on
producing geometric embeddings of graphs
satisfying various aesthetic constraints.
After the geometric embedding is specified, there is an additional step 
that is often overlooked or ignored: 
assigning display colors to the graph's vertices.
We study
the additional aesthetic criterion of assigning distinct colors to
vertices of a geometric graph
so that the colors assigned to adjacent vertices are as
different from one another as possible.
We formulate this as a problem involving perceptual metrics in color
space and we develop algorithms for solving this problem by
embedding the graph in color space.
We also present an application of this work to a distributed load-balancing
visualization problem.

\end{abstract}

\section{Introduction}
Graphs are frequently visualized by embedding them in geometric spaces.
That is, geometric representations are natural tools for
visualizations; hence, we embed graphs in geometric spaces in order to
display them.
For instance, producing geometric embeddings of combinatorial graphs
so as to satisfy various aesthetic constraints is a major component of
\emph{graph drawing}
(e.g., see~\cite{dett-gd-99,jm-gds-03,kw-dgmm-01,n-pgd-04}).

Once a graph has been embedded in a geometric space, such as $\R^2$,
we refer to it as a \emph{geometric graph}.
That is, a geometric graph
is a graph $G=(V,E)$ such that 
the vertices are geometric objects in $\R^d$ and the edges are
geometric objects connecting pairs of vertices.
Note that this definition is more general
than the definition of ``{geometric graph}'' popularized by 
Alon and Erd{\"o}s~\cite{ae-degg-89}, in that
they define a geometric graph to be a graph $G=(V,E)$ such that the
vertices are distinct points in $\R^2$ and edges are straight
line segments. 
% There is a developing interest 
% in such geometric graphs
% (e.g., see~\cite{f-gga-04,p-ggt-99,p-ggt-04,p-ttgg-04,t-cagg-98}).
%as well as a rich historical heritage for them
%(e.g., see~\cite{f-slrpg-48,k-cdf-63,t-crg-60,t-hdg-63}).
% Nevertheless, inspired by work in
% the graph drawing literature
% (e.g., see~\cite{dett-gd-99,jm-gds-03,kw-dgmm-01,n-pgd-04}),
% we take the more general approach.
For example, we allow a geometric graph 
to be a planar map, where the vertices
are regions and the edges are defined by regions that share a common
border.

Intuitively, a geometric graph $G$ is a graph that is ``almost
drawn,'' because displaying $G$ requires assigning
colors to its vertices.
One obvious method of doing this---a very common one---is to ignore
the issue and color all the vertices black.  In this paper, we 
examine the color-choosing step more carefully.
In particular, are interested in methods for choosing colors for
the vertices of a geometric graph so as to make distinctions
between vertices as apparent as possible.
% Unlike the traditional graph coloring problem, we are not
% interested in using the fewest colors possible, since any 24-bit
% color display can render 16.2 million colors.
We are also interested in the related \emph{map coloring} problem, 
where we color the
faces of a map so as to make the distinctions between adjacent faces
as strong as possible.
Part of the challenge is choosing a good set of colors, but 
we also want to assign
colors to vertices in a way that makes the colors assigned to
adjacent vertices as different as possible.
That is, we are interested in a bi-criterion color
assignment problem, where all the colors are different from one another and
adjacent colors are really different.

\subsection{Previous Related Work}
Graph coloring is a classic problem in algorithmic graph theory
(e.g., see~\cite{bm-gta-76}).
Given a graph $G$, the traditional version of this 
problem is to color the vertices of $G$ with as few
colors as possible so that adjacent vertices always have different colors.
The traditional 
graph coloring problem is posed as a ``coloring'' problem purely 
for abstraction's sake, however: no paint or pixels are involved. 
Even so, there has been some prior work on algorithms for coloring
geometric graphs (in the traditional sense).
For example, there has been some prior research on coloring
quadtrees~\cite{BerEppHut-Algo-02}, intersection
graphs~\cite{Epp-SODA-04-bgig}, and arrangements~\cite{FelHurNoy-SODA-00}.
In addition, there has been a host of prior work
on the traditional version of graph coloring
for purely combinatorial graphs
(e.g., see~\cite{bm-gta-76}).

Also of interest is
work that has been published in the information visualization 
literature on methods for choosing colors effectively for data 
presentation.
Healey~\cite{healey96choosing} presents a heuristic for choosing a
well-separated set of colors for visualizing segmentation data in images.
Likewise, Levkowitz and Herman~\cite{lh-clid-92},
Robertson~\cite{r-vcg-88},
and Ware~\cite{w-csum-88}
discuss various ways for effectively building color maps that correspond to 
data values in an image or data visualization (e.g., a bar chart histogram).
Rheingans and Tebbs~\cite{rt-tdecm-90}
describe an interactive approach that
constructs a color scale by 
tracing a path through color space.
Brewer~\cite{b-cugdr-99} describes several guidelines for choosing
colors for data visualization, focusing primarily on ways of
representing linear numerical scales.
There are also several good books on the subject of
color use for data visualization 
(e.g., see~\cite{s-fgdc-03,t-vdqi-83,t-ei-90}).
In spite of this wealth of previous work on color selection for data
visualization, we are unfamiliar 
with any prior work that uses adjacency information to select
dissimilar colors for visualization purposes.

%-------------------------------------------------------------
\subsection{Our Results}
In this paper, we investigate the following 
problem for geometric graph coloring:
\begin{quote}
\emph{Maximizing minimum color difference}.
Given a geometric graph $G$ and a color
space $\cal C$, assign visibly distinct
colors from $\cal C$ to the vertices of
$G$ so to maximize the minimum color difference across the endpoints
of edges in $G$.
\end{quote}
%
% SAVE THIS ONE FOR LATER - WE HAVE ENOUGH FOR NOW - DE
%
%\item
%\emph{Maximizing weighted minimum color difference}.
%Suppose we are again given a geometric graph $G$ and 
%color space $\cal C$, but instead of wishing to 
%maximize the minimum color difference across
%each edge of $G$, we instead choose to
%maximize a weighted difference.
%For example, the weight of an edge could be equal to its length (for
%a primal planar embedding) or the length of a face boundary in a
%planar dual (for a dual embedding). 
%In the primal version of geometric graph coloring, for example, 
%adjacent vertices
%that are close together would get very different colors.
%\end{itemize}
%
We investigate this problem in terms of embeddings of $G$ in the 
human-perceptible subset of the color space $\cal C$. 
This embedding of $G$ in $\cal C$ is purely to find good
colors to assign to the vertices of $G$, however. The actual
coordinates for $G$'s vertices and equations for $G$'s edges
will still use $G$'s geometric embedding in $\R^d$ (e.g., as
produced by an existing graph-drawing algorithm).
Nevertheless, the placement of vertices and edges in our embedding of
$G$ in $\cal C$ implies a ``goodness'' score on the degree to which
adjacent vertices are well-separated and non-adjacent vertices are
fairly-separated (which corresponds to a similar degree of separation for
the vertex colors when we display $G$ using its original geometry).
We design a force-directed algorithm to produce such embeddings.
%, with
%uniform edge weights in the unweighted case and
%multiplicatively-weighted edges in the weighted case.

By planar duality,
our algorithms are also applicable to the \emph{map coloring}
problem, where we are given a planar map and asked to color the
regions with distinct colors so that the color difference between
bordering regions is as large as possible.
% (possibly weighted by the length of the common border).
We give an application of this map coloring problem to an interesting
data visualization problem for load-balancing distributed numerical
algorithms.

% We feel that this idea of coloring graphs to optimize the visual 
% distinctiveness of nearby vertices is likely to lead 
% to additional interesting research problems, as well.

\section{Color Space}
Since we wish to assign colors to the
vertices of a geometric graph
so that colors assigned to adjacent vertices look as different 
as possible, it is useful to have a precise, mathematical
notion of color and color difference.

Pure colors can be defined in terms of wavelengths of light, with the
visible spectrum of colors going roughly from 400 nm (violet) to 800 nm (red).
Humans perceive color, however, as a combination of 
intensity signals from three types of cone cells in our eyes:
\begin{itemize}
\item
\textbf{S cone cells}: These cells respond to short wavelengths and
typically have their peak transmission around 440 nm (violet). 
(For historical reasons, these cells are often referred 
to as ``blue'' cone cells.)
\item
\textbf{M cone cells}: These cells respond to medium wavelengths and
typically have their peak transmission around 550 nm (yellow-green). 
(For historical reasons, these cells are often referred
to as ``green'' cone cells.)
\item
\textbf{L cone cells}: These cells respond to long wavelengths and
typically have their peak transmission around 570 nm (yellow). 
(For historical reasons, these cells are often referred
to as ``red'' cone cells.)
\end{itemize}

This physiology forms the basis of all color displays, from
old-fashioned color TVs to modern-day color LCD panels and 
plasma displays, for these displays create what we perceive as colors
by an additive combination of three color intensities, such as red, green, and
blue (RGB, as exemplified by the sRGB space~\cite{sacm-srgb-96} 
%and Adobe RGB color spaces
used by many digital cameras and color displays).
This physiology also forms the 
basis of most color printing, as well, where printers create what we
perceive as colors by the subtractive combination of three color
intensities, such as cyan, magenta, and yellow (CMY).
Thus, colors can be viewed as belonging to a three-dimensional
\emph{color space}.
Moreover, RGB color spaces, which define three-dimensional cubes of color
values, correspond to 
the way most modern devices display colors.

\begin{figure}[b!]
\centering
\includegraphics[height=2.45in]{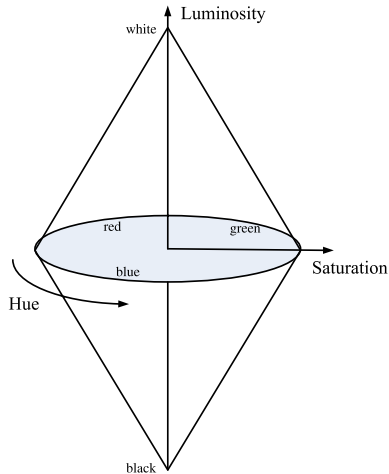}
\caption{The Hue-Saturation-Luminosity (HSL) color space.}
\label{fig-hls}
\end{figure}

Ironically, even though RGB spaces are the most popular for display devices, 
humans are very poor at interpreting the perceived color that
results from the addition of intensity values in red, green, and blue.
Moreover, perceived color differences do not define a uniform
metric in RGB spaces.
Our brains instead use the following notions:
\begin{itemize}
\item \textbf{Hue}: the actual color, e.g., ``blue,''
``yellow,'' ``orange,'' etc., as defined by a radial value around a
color wheel.
\item \textbf{Saturation}: the vividness or dullness of the color.
\item \textbf{Luminosity}: the lightness or darkness of the color.
\end{itemize}
Thus, human perceived color defines a three-dimensional color space,
called HSL, which corresponds to two cylindrical cones joined at
their base, as shown in Figure~\ref{fig-hls}.
The two apexes of these cones correspond to opposite corners in RGB
space.
As with RGB, however, perceived color differences do not define a
uniform metric in the HSL color space.
Moreover, the geometry of the HSL space makes choosing colors inside
its double-cone of visible colors more challenging.

CIE L*a*b* (or ``Lab,'' for short) is an absolute color
space that defines each color uniquely as a combination of Luminosity
(L), a value, a*, which is a signed number that indicates the degree of
magenta (positive) or green (negative) in a color, and a value, b*,
which is a signed number that indicates the degree of yellow
(positive) or blue (negative) in a color.
Geometrically,
Lab is a slightly distorted version of the HSL double-cone,
with color points addressed using Cartesian coordinates.
Thus, defining the subset of visible colors in Lab space is admittedly more
challenging.
Offsetting this drawback, however, is the fact
that empirical evidence supports the claim that
Euclidean distance
in this color space corresponds to perceptual color 
difference~\cite{rptb-eedmc-01}.
There is a related, CIE L*u*v* color space, which also is
designed to provide a uniform color-difference metric, but the Lab color space
seems to be more uniform.
Thus, the Lab color space is the more popular of the two.

There is a tradeoff between
the two most popular color spaces, then.
RGB corresponds better to display hardware and it defines a
simple cube geometry for the space of visible colors. But perceived
color difference is not a uniform metric in RGB.
Lab space, on the other hand, has a more complex geometry and
requires a translation to RGB for display purposes, but it supports a
uniform color-difference metric.
In this paper, therefore, we explore color choosing algorithms for
both of these spaces.

\input{application}

\section{What is a Good Coloring?}

Formally,
our problem can be stated as follows. We are given as input an undirected graph $G$, the vertices of which have been partitioned into \emph{regions} $r_i$. We would like to display this region structure, overlaid on a conventional drawing of the graph, by assigning distinct colors to vertices in different regions. Our task is to choose a color for each region, satisfying the following constraints:

\begin{itemize}
\item Each region must have a different color, and
the colors assigned to regions must be visually distinct.

\item If two regions $r_i$ and $r_j$ are adjacent in $G$ (that is, if some vertex in $r_i$ and some vertex in $r_j$ are adjacent), then it is especially important that regions $r_i$ and $r_j$ be given dissimilar colors. We desire that the colors of such adjacent regions be as dissimilar as possible, subject to the first constraint that all region colors be visually distinct.
\end{itemize}

To solve this problem, we construct a \emph{region graph} $R$ (as in
Figure~\ref{fig:partition}(b)). We form one vertex in $R$ per region
$r_i$, with regions $r_i$ and $r_j$ connected by an edge in $R$ if
and only if they are adjacent.  We view the problem of
assigning colors to the regions as one of embedding $R$ geometrically,
into a three-dimensional space representing the gamut of colors
available on the display device. Ideally, distances in this space
should represent the visual dissimilarity of a pair of colors. 
As mentioned above, color
spaces such as Lab have been designed so that this dissimilarity can be
approximated by a Euclidean distance in that space.

Thus, we have a geometric graph embedding problem: assign color
coordinates in a color space $\cal C$
to each vertex of the region graph $R$, according to the
dissimilarity criteria identified above.  However, unlike the embedding
problems coming from traditional graph drawing problems, we want to place vertices
so that edges are long rather than short.

In order to formalize the problem, we define a {\em coloring} to be any mapping $\chi$ from the vertices of $R$ to the color space of interest.  
Let $d_{i,j}$ denote the distance between $\chi(r_i)$ and $\chi(r_j)$, as measured by an appropriate distance function corresponding to visual dissimilarity.
Let $D$ be the dimension of the color space; in most instances we will have $D=3$.
Let $n$ be the number of vertices in the region graph.
For any region $r_i$, let $N_i$ denote the set of adjacent regions in $R$.
Finally, let $\Delta$ denote the diameter of the color space into which we are embedding our region graph.
We define a quality measure $q(\chi)$ by the following equation:
$$
q(\chi)=\sum_{r_i}( 
\sum_{r_j\in R\setminus\{r_i\}} \frac{1}{d_{i,j}^{D+1}}
+
\frac{n^{1+1/D}}{\Delta^D}
\sum_{r_j\in N_i} \frac{1}{d_{i,j} |N_i|}
).
$$
One of our goals in defining a function of this form is that, by making the quality a sum of relatively simple terms, we may find its gradient easily, simplifying the application of standard numerical optimization techniques.
There are two terms per region in this sum, both normalized to be of roughly equal significance.

The first term has the form
$\sum_{r_j} d_{i,j}^{-(D+1)}$.
We expect, in a good embedding of the region graph, that the regions will be roughly uniformly distributed around the region graph. The exponent $D+1$ in this term is chosen with this assumption in mind:
for infinitely many uniformly spaced regions with a spacing of $\delta$,
$\sum d_{i,j}^{-(D+1)}$ will converge to $\Theta(\delta^{-(D+1)})$, being influenced most strongly by the regions nearest $r_i$. On the other hand, a similar sum with an exponent of $D$ or less would diverge, and thus lowering the exponent in this term would cause our quality measure to be dominated more by global than local concerns.
For $n$ vertices in a $D$-dimensional region of diameter $\Delta$,
we expect spacing $\delta=\Theta(\Delta n^{-1/D})$,
and thus we expect
$$\sum_{r_j\in R\setminus R_i} \frac{1}{d_{i,j}^{D+1}}
=\Theta(\delta^{-(D+1)})
=\Theta(\frac{n^{1+1/D}}{\Delta^{D+1}})
.$$

The second term has the form
$\sum_{r_j\in N_i} 1/(d_{i,j} |N_i|)$.
We hope, especially in the case of relatively sparse region graphs, to have $d_{i,j}$ roughly proportional to $\Delta$ for all edges between adjacent regions $r_i$ and $r_j$. If these edges are all sufficiently long, the normalization by $|N_i|$ will leave this term roughly proportional to $1/\Delta$.
The low exponent on the distance is acceptable as we wish this part of the quality measure to act long-range, causing adjacent regions to be placed far apart.
The normalization factor
$n^{1+1/D}\Delta^-D$
prior to the second term in our definition of $q$ is chosen to make the two terms of the sum roughly proportional.

\section{Finding a Good Coloring}

The problem of finding a good coloring can be approached with a standard gradient descent or hill climbing heuristic: choose initial vertex locations in color space, and then gradually move the locations in a direction that causes the most local improvement in our quality measure.
This requires calculating the gradient of our quality measure,
which is most easily done when our color space forms a normed vector space, 
preferably Euclidean. 
Lab color is ideal for this task, as it has been designed so that Euclidean distances in Lab color closely approximate visual dissimilarity. 
The same approach can also be applied directly to RGB-based color spaces such as sRGB, with some degradation in the goodness of fit between our quality measure and the visual dissimilarity of the resulting colors.

As our quality measure is linear, we may compute the gradient separately for the term of it applying to each region $r_i$. 
The gradient at $r_i$ is a vector-valued quantity, formed by summing for each $r_j$ a vector directed away from $r_j$. 
If $r_i$ and $r_j$ are not adjacent, this vector has length $(D+1)/d_{i,j}^{D+2}$. 
If $r_i$ and $r_j$ are adjacent, we add another vector in the same direction with length
$$
\frac{n^{1+1/D}}{|N_i|\Delta^D}d_{i,j}^{-2}.
$$

Gradient descent with these vectors will cause the
locations of regions in color space to spread apart rapidly. 
But we do not allow this to continue unconstrained, 
as we must confine the colors of each region to
the gamut of displayable colors on the intended output device.  We
considered several options for this confinement: 

\begin{itemize}
\item 
We could add an additional term in the quality measure penalizing
colors outside the allowable gamut. However, we do not wish to penalize
colors near the boundaries of the gamut, because those boundaries
provide saturated colors that are easy to visually distinguish. Nor do
we wish to allow colors to drift very far beyond the gamut. So the
penalty term would have to have a very steep derivative, making the
numerical optimization more difficult.
\item 
We considered clipping any color outside the gamut to the
nearest color within the gamut. Like the penalty term, this method
would affect only the colors that reach the boundaries of the allowable
gamut, and the numerical optimization procedure would have difficulty
propagating the effects of this clipping to the interior of the gamut.
More significantly, this truncation could distort points
near boundaries of the color space where the tangent plane is not 
perpendicular to the line from the point to the center of the gamut. 
Effectively, the truncation and the outward repulsive forces of the
gradient descent would push 
these points along the boundary away from the center.
\item 
We experimented with a procedure that, after each step of gradient
descent, rescales the entire color space, so that all vertices again
lie within the gamut of allowable colors. This seems to work
acceptably well for symmetric color spaces such as the sRGB gamut.
However, when we tried it with Lab colors, for which the color space is
more stretched out in some directions than others, we found that this
method tended to accentuated this stretching, causing the gamut to be
compressed in the other directions. In particular, this led to
significant desaturation of the resulting Lab colors.
\item 
We finally settled on the following procedure: after each step of gradient
descent, rescale (rather than truncating) the out-of-gamut points,
while leaving the other points in place. In our experiments this method
performed better than the other ones above, allowing the gradient
descent to improve the color placement without distorting the gamut.
\end{itemize}

Our implementation chooses initial vertex locations at random within the color gamut. 
Then in each iteration it attempts to move the locations of the vertices in 
color space, one vertex at a time by three types of moves: random jumps, swaps 
with other vertices, and moving by a fixed step size in the gradient direction. 
For each of these three move types our algorithm accepts the move only when 
it improves the overall quality of the coloring. 
If an iteration fails to find any quality improvement, we reduce the step size 
and terminate the algorithm when this step size falls below a preset threshold.

\section{Results of Our Implementation}

\begin{figure}[t]
\centering
\includegraphics[width=2in]{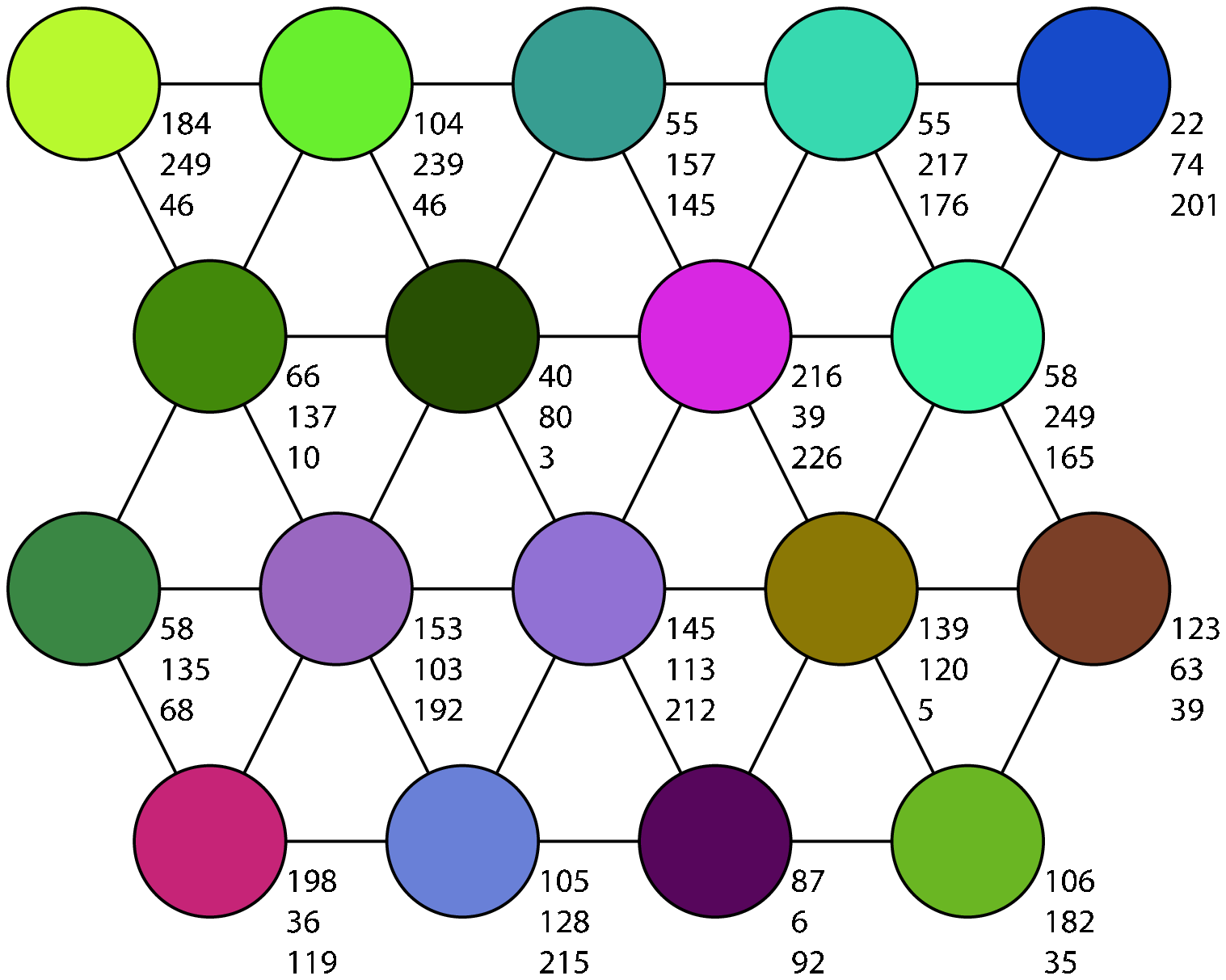}\qquad
\includegraphics[width=2in]{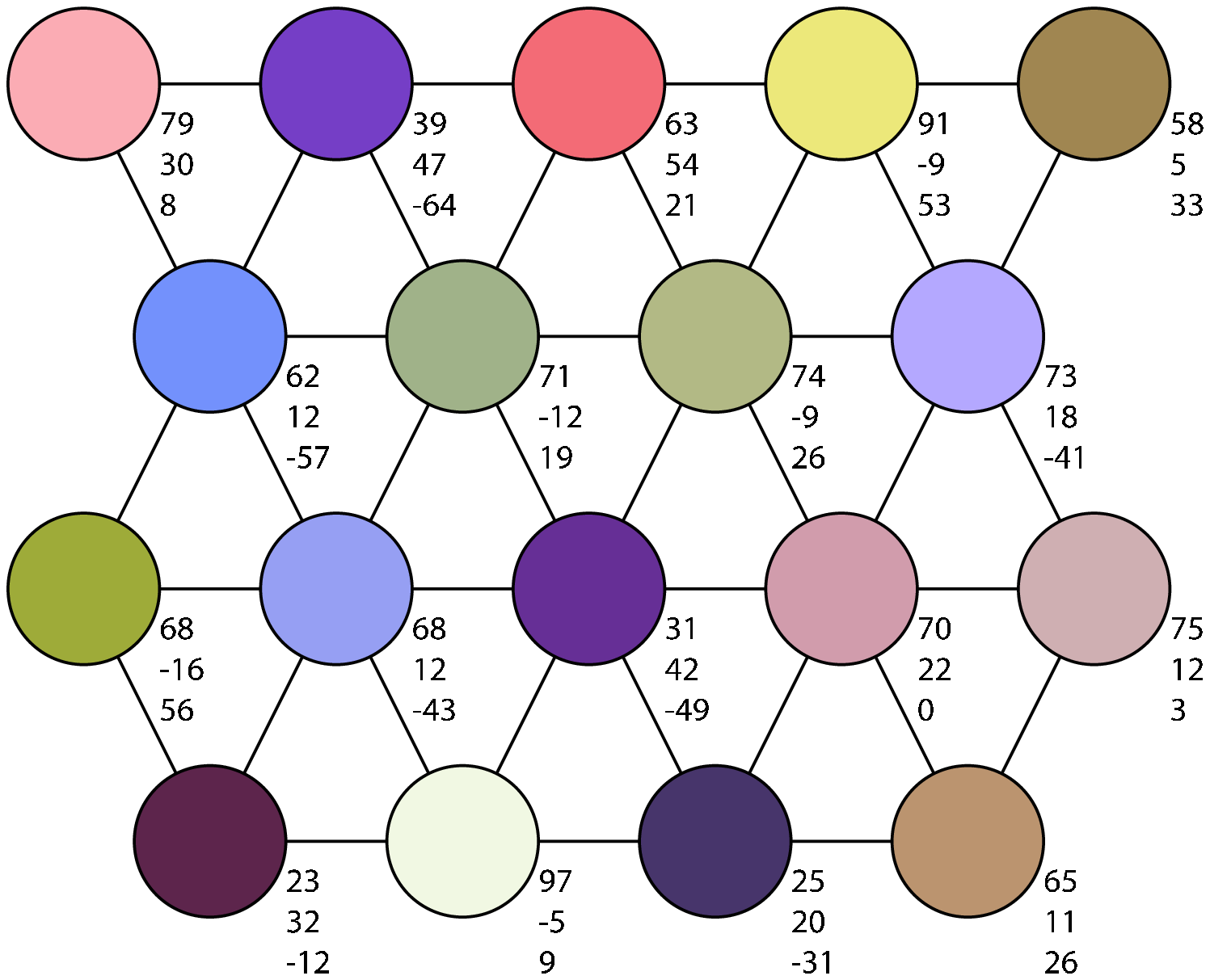}
\vspace*{-10pt}
\caption{Results of implementation: random assignment of colors to vertices. Left: random sRGB colors; right: random Lab colors.}
\label{fig:random}
\end{figure}

As a proof of concept, we implemented our algorithm both for the sRGB
and Lab color spaces, and compared the results with those from an
algorithm that chooses colors randomly.  As the color spaces we use form eight-vertex convex polyhedra, our algorithm will tend to choose colors at those vertices
for graphs with eight or fewer regions. For this reason, we chose for
our experiments a larger region graph
in the form of an eighteen-vertex triangulation.

We believe that, ultimately, the most appropriate way to evaluate our results is human usability testing, but such experiments are beyond the scope of this paper. Our numerical quality measure is not suitable for comparing different algorithms, first because it is specific to a color space and would not allow easy comparison of colorings in different spaces, and second, because any such comparison would not test how well our quality measure itself models the ease of understanding of drawings using our colorings. 
Thus, we only attempted a limited subjective analysis, based due to space limitations only on the results of a single run of each algorithm. Once each of our algorithms was working correctly, we ran it only once in order to avoid biasing our results by choosing subjectively among multiple runs; we note that an automated choice among multiple runs based on our quality measure would be possible, but would not differ in principle from a single run of a more sophisticated optimization procedure than our randomized hill climbing algorithm. 
The random nature of our algorithms means that the precise colors generated in our evaluation are not repeatable, and more systematic usability testing is needed to verify our results.

In the first implementation (Figure~\ref{fig:random}(left)), we chose random
colors independently for all vertices, uniformly among the
$2^{24}$ possible sRGB values. As expected, this did not work very
well. The random assignment did not prevent several very similar colors
from being chosen, often for adjacent regions. We performed a similar experiment with colors chosen uniformly at random in Lab color space, within the convex hull of the eight extreme sRGB colors (Figure~\ref{fig:random}(right)); the resulting colors seemed less heavily dominated by greens than the random sRGB results, but still included several very similar adjacent pairs of colors.

\begin{figure}[t]
\centering
\includegraphics[width=2in]{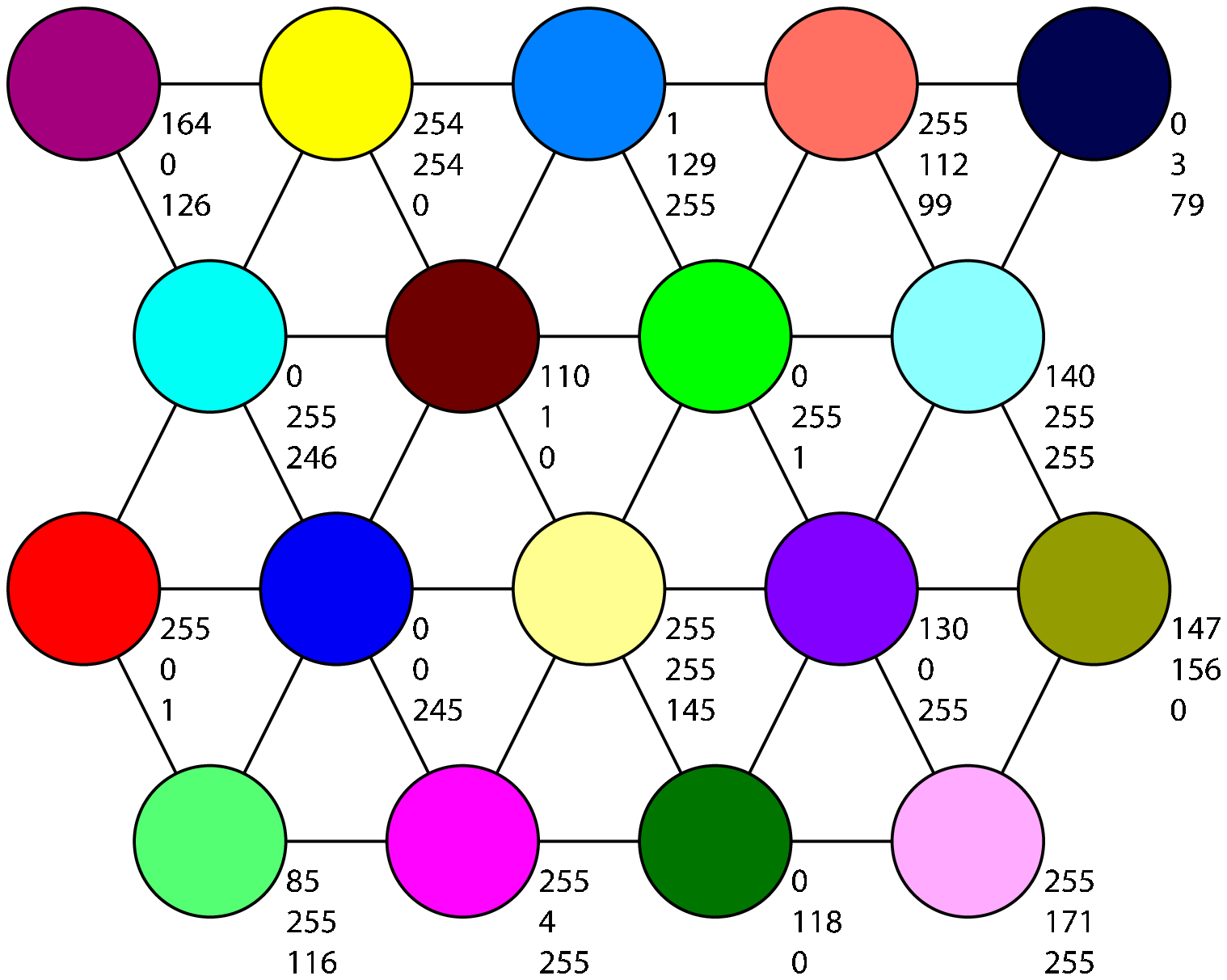}\qquad
\includegraphics[width=2in]{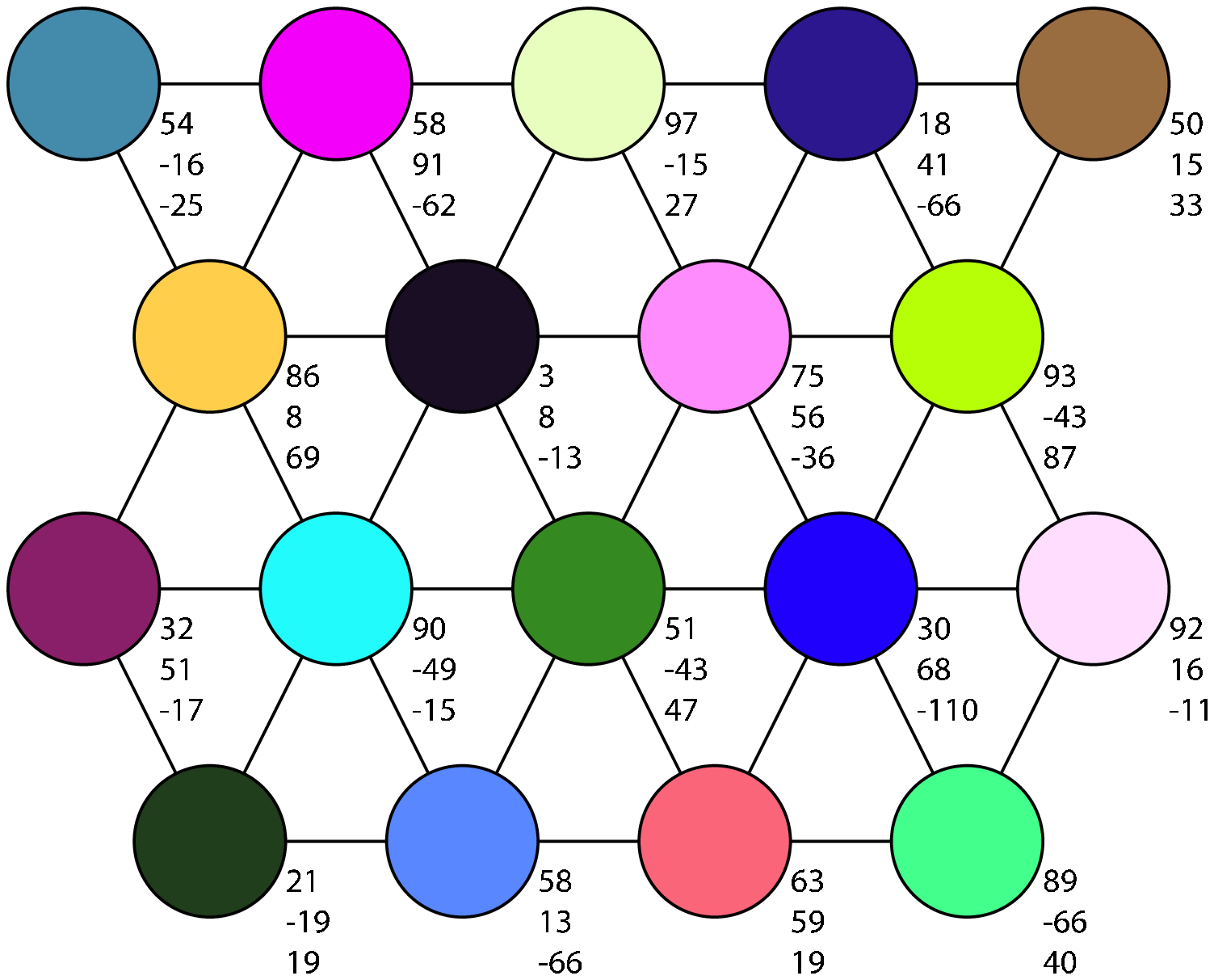}
\vspace*{-10pt}
\caption{Results of implementation: gradient descent in sRGB color space (left) and in Lab color space (right).}
\label{fig:descent}
\end{figure}

In the second implementation (Figure~\ref{fig:descent}(left)) we applied our gradient descent optimization algorithm directly to the sRGB color space, using the quality measure we defined earlier via the Euclidean distance in this space despite the fact that this distance is known to fit human vision poorly. 
The algorithm chose a diverse selection of well saturated colors, and all pairs of adjacent regions have easily distinguishable colors. However, there are several nonadjacent colors that are difficult to distinguish: two yellows (254,254,0 and 255,255,145), two cyans (0,255,246 and 140,255,255), three blues (1,129,255, 0,0,245, and 130,0,255), two reds (255,112,99 and 255,0,1), and two pinks (255,4,255 and 255,171,255). We believe these faults are due to the poor match between Euclidean distance in sRGB color space and human visual dissimilarity.

Finally, we applied our gradient descent algorithm for coordinates in the Lab color space (Figure~\ref{fig:descent}(right)).
The gamut of representable colors in Lab space is significantly larger than that for sRGB, so in order to ensure that our algorithm generated colors that could be displayed, we restricted all colors to a gamut formed geometrically as the convex hull in Lab space of the eight colors black, white, red, green, blue, cyan, magenta, and yellow forming the most extreme values of the sRGB color space.
When our gradient descent algorithm caused vertices to be assigned colors 
outside this convex hull, as described above, we returned them to the gamut 
by a rescaling operation centered at the neutral gray color with Lab 
coordinates $50,0,0$.
As with the sRGB output, the result of this algorithm was a collection of diverse well saturated colors, with all pairs of adjacent regions having easily distinguishable colors. 
Compared to the sRGB results, there were fewer sets of difficult to distinguish nonadjacent colors: primarily the two darker pinks (58,91,-62 and 75,56,-36). It is also somewhat difficult to distinguish the dark green (21,-19,19) from the black (3,8,-13).
On the whole, it seems that using Lab color has led to a better selection of colors, and equally good assignment of the chosen colors to the vertices of the region graph.

\section{Conclusion}
We have given what we believe is the first color
assignment algorithm that uses adjacency information in the input
geometric graph to choose colors that are very different for
adjacent vertices.
% This approach therefore can be viewed as a compromise between traditional
% color assignment, which does not take adjacency information into
% consideration, and traditional graph coloring, which only considers
% adjacency information.
% 
For possible future work, one could consider a weighted version of
the problem, where edges of the
input geometric graph are weighted (e.g., by length) and we wish to
assign colors so that the colors assigned to vertices of
low-weight edges are more dissimilar than those on high-weight edges.
Another interesting adaptation would be to perform our
color assignment algorithm for color spaces corresponding to 
color-blind people (of which there are six types that collectively
make up roughly 8\% of the male population).

\small\raggedright
\bibliographystyle{abbrv}
\bibliography{geom,geom-updates,goodrich,extra-de,conf,extra}
%
% The above bib files are available in ~goodrich/lib/bibtex
%

\end{document}

%% file: application.tex
\section{Application}

A specific application motivating this research is
a problem in distributed programming.
The Navigational Programming (NavP) methodology \cite{pan04} for converting
a sequential program into a parallel distributed program using
migrating threads consists of the following three steps:
\begin{enumerate}
\item {\bf Data Distribution:\/}
 The data used by the program is distributed over the network.
The guiding heuristic principle is minimizing communication cost while
balancing the load on each processing element (PE).
\item {\bf Computation Distribution:\/}
 Navigational commands (``hop'' statements) are inserted into the
sequential code.  
This step produces
a {\it distributed
sequential program\/}, a single thread that ``follows'' the data
through the network.
\item  {\bf Pipelining:\/}
The single migrating thread produced in step~2 is broken into multiple
threads, which are then formed into a pipeline by adding appropriate
synchronization commands.
\end{enumerate}
The methodology incorporates a feedback loop: information obtained in step~3
can be used to improve the data distribution in a subsequent application
of the three steps.

The data distribution step is based on constructing a
Navigational Trace Graph (NTG), which relates communication costs
to data placement, and then applying a graph partitioning heuristic.
In the NTG, the vertices are the data elements, and edge weights
between vertices reflect the cost of placing the corresponding
data elements on different machines~\cite{pan06}.
Among the factors influencing the edge weights between two data items are
(1) whether one of them is used directly in the computation
of the other; (2) whether they are physically allocated close together in the
sequential program; (3) whether they are referenced in temporally consecutive
statements in the sequential program.  The first factor is a source of
communication overhead if the data elements are assigned to different PE's,
while the second and the third factors capture locality information that
is implicit in the sequential code and may affect performance
(e.g., by increasing cache reuse).
Additional factors affect the partitioning: these include balancing the
computational load and amount of data on each PE, and introducing
constraints that certain data elements must be on different PE's
(so as not to preemptively exclude parallelism that might be introduced in
step 3).

Since the interaction of these constraints can be complex,
it is important to be able to visualize the resulting data
partitions.  One ingredient of a good visualization tool is effective
use of color.  
Because the individual sets in the data partition are not necessarily connected,
the color-assignment scheme should follow
two basic principles: 
(1) The colors assigned to regions in the partition should be
highly dissimilar, to make it easy to see
the boundaries between regions.
(2) All colors used should be somewhat dissimilar from each other, so that it
is apparent which disconnected regions belong to the same set in the partition.

\begin{figure}[t]
\centering
\begin{tabular}{c@{\hspace*{0.5in}}c}
\includegraphics[height=1.75in]{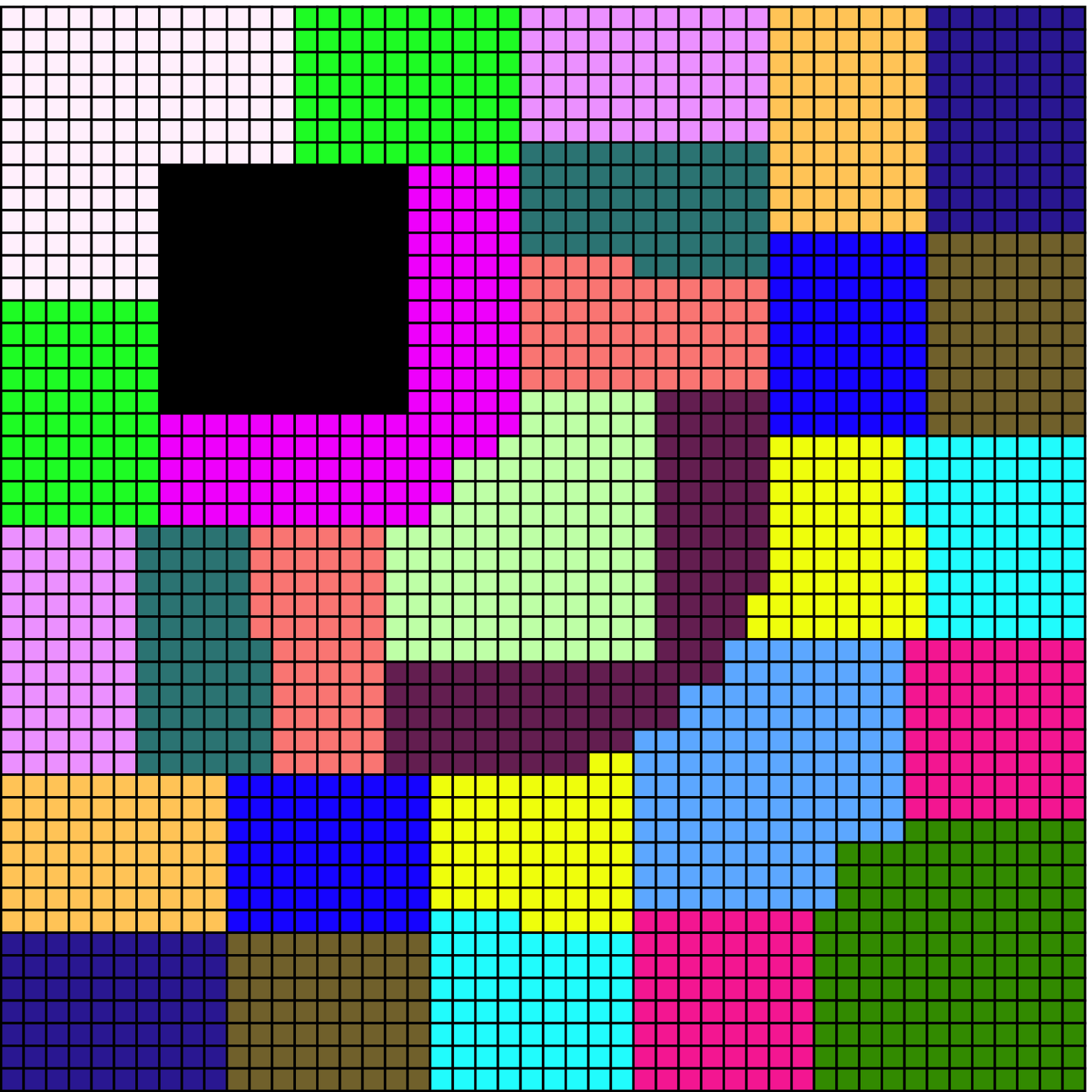}
&
\includegraphics[height=1.75in]{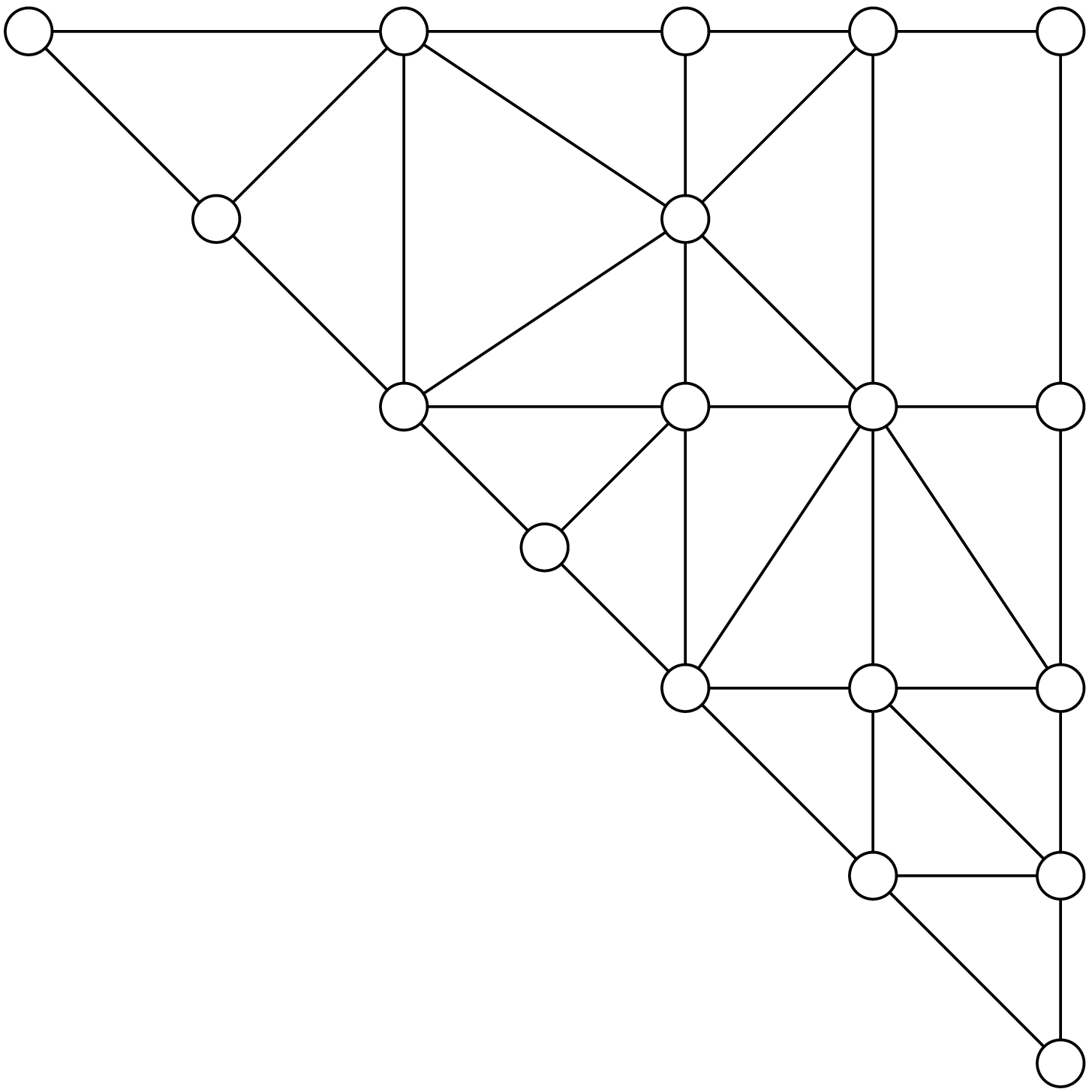} \\
(a) & (b)
\end{tabular}
\caption{The Navigational Programming application:
(a) an example partition of an array of data into 18 regions for 
a NavP application, colored by our algorithm; and 
(b) the corresponding graph of adjacent regions.}
\label{fig:partition}
\end{figure}

An example of a partition produced by this system is shown in Figure~\ref{fig:partition}.  The underlying sequential
code for which this partition was constructed, adapted from Lee and Kedem~\cite{lee02}, can be conceptualized as a sequence
of scans over a square matrix, where the scans alternate between row-major
and column-major order and also alternate directions.  In each scan, the
value of each element $A[i,j]$ is computed as a function of its neighbors
and also the neighbors of its transposed element $A[j,i]$.  As can be
seen, the partition is somewhat irregular.  Because of the strong data
affinity between each element and its transpose, the sets are symmetric 
about
the diagonal of the matrix, and some of them are not connected.
Thus, because of the irregularity of the partitioning and the
potential complexity of the partitions, it is useful to have a high-quality assignment of colors to regions
in order to visualize the regions and their interactions.

%% file: paper.bbl
\begin{thebibliography}{10}

\bibitem{ae-degg-89}
N.~Alon and P.~Erd{\H o}s.
\newblock Disjoint edges in geometric graphs.
\newblock {\em Discrete Comput. Geom.}, 4:287--290, 1989.

\bibitem{BerEppHut-Algo-02}
M.~W. Bern, D.~Eppstein, and B.~Hutchings.
\newblock {Algorithms for coloring quadtrees}.
\newblock {\em Algorithmica}, 32(1):87--94, 2002.

\bibitem{bm-gta-76}
J.~A. Bondy and U.~S.~R. Murty.
\newblock {\em Graph Theory with Applications}.
\newblock Macmillan, London, 1976.

\bibitem{b-cugdr-99}
C.~A. Brewer.
\newblock Color use guidelines for data representation.
\newblock In {\em Proc. Section on Statistical Graphics, American Statistical
  Association}, pages 55--60, 1999.

\bibitem{dett-gd-99}
G.~{Di Battista}, P.~Eades, R.~Tamassia, and I.~G. Tollis.
\newblock {\em Graph Drawing}.
\newblock Prentice Hall, Upper Saddle River, NJ, 1999.

\bibitem{Epp-SODA-04-bgig}
D.~Eppstein.
\newblock {Testing bipartiteness of geometric intersection graphs}.
\newblock In {\em Proc. 15th Symp. Discrete Algorithms}, pages 853--861. ACM
  and SIAM, 2004.

\bibitem{FelHurNoy-SODA-00}
S.~Felsner, F.~Hurtado, M.~Noy, and I.~Streinu.
\newblock {Hamiltonicity and colorings of arrangement graphs}.
\newblock In {\em Proc. 11th Symp. Discrete Algorithms}, pages 155--164. ACM
  and SIAM, 2000.

\bibitem{healey96choosing}
C.~G. Healey.
\newblock Choosing effective colours for data visualization.
\newblock In R.~Yagel and G.~M. Nielson, editors, {\em {IEEE} Visualization
  '96}, pages 263--270, 1996.

\bibitem{jm-gds-03}
M.~J\"unger and P.~Mutzel.
\newblock {\em Graph Drawing Software}.
\newblock Springer, 2003.

\bibitem{kw-dgmm-01}
M.~Kaufmann and D.~Wagner.
\newblock {\em Drawing Graphs: {Methods} and Models}, volume 2025 of {\em
  Lecture Notes in Computer Science}.
\newblock Springer-Verlag, 2001.

\bibitem{lee02}
P.~Lee and Z.~M. Kedem.
\newblock Automatic data and computation decomposition on distributed memory
  parallel computers.
\newblock {\em ACM Trans. Programming Languages and Systems}, 24(1):1--50,
  2002.

\bibitem{lh-clid-92}
H.~Levkowitz and G.~T. Herman.
\newblock Color scales for image data.
\newblock {\em IEEE Computer Graphics and Applications}, 12(1):72--80, 1992.

\bibitem{n-pgd-04}
T.~Nishizeki.
\newblock {\em Planar Graph Drawing}, volume~12 of {\em LNSC}.
\newblock World Scientific, 2004.

\bibitem{pan04}
L.~Pan, M.~K. Lai, K.~Noguchi, J.~J. Huseynov, L.~F. Bic, and M.~B.
  Dillencourt.
\newblock {Distributed parallel computing using Navigational Programming}.
\newblock {\em Int. J. Parallel Programming}, 32(1):1--37, 2004.

\bibitem{pan06}
L.~Pan, J.~Xue, M.~K. Lai, L.~Bic, M.~B. Dillencourt, and L.~Bic.
\newblock {Toward automatic data distributions for migrating computations}.
\newblock Submitted for publication, 2006.

\bibitem{rt-tdecm-90}
P.~Rheingans and B.~Tebbs.
\newblock A tool for dynamic explorations of color mappings.
\newblock {\em Computer Graphics}, 24(2):145--146, 1990.

\bibitem{r-vcg-88}
P.~K. Robertson.
\newblock Visualizing color gamuts: {A} user interface for the effective use of
  perceptual color spaces in data displays.
\newblock {\em IEEE Computer Graphics and Applications}, 8(5):50--64, 1988.

\bibitem{rptb-eedmc-01}
Y.~Rubner, J.~Puzicha, C.~Tomasi, and J.~M. Buhmann.
\newblock Empirical evaluation of dissimilarity measures for color and texture.
\newblock {\em Computer Vision and Image Understanding}, 84:25--43, 2001.

\bibitem{sacm-srgb-96}
M.~Stokes, M.~Anderson, S.~Chandrasekar, and R.~Motta.
\newblock {A Standard Default Color Space for the Internet -- sRGB}.
\newblock Available online at
  http://www.w3.org/pub/WWW/Graphics/Color/sRGB.html, 1996.

\bibitem{s-fgdc-03}
M.~Stone.
\newblock {\em A Field Guide to Digital Color}.
\newblock Morgan Kaufmann, 2nd edition, 2003.

\bibitem{t-vdqi-83}
E.~R. Tufte.
\newblock {\em The Visual Display of Quantative Information}.
\newblock Graphics Press, Cheshire, Connecticut, 1983.

\bibitem{t-ei-90}
E.~R. Tufte.
\newblock {\em Envisioning Information}.
\newblock Graphics Press, Cheshire, Connecticut, 1990.

\bibitem{w-csum-88}
C.~Ware.
\newblock Color sequences for univariate maps: {Theory}, experiments, and
  principles.
\newblock {\em IEEE Computer Graphics and Applications}, 8(5):41--49, 1988.

\end{thebibliography}
